# Chemical and Structural Disorder in Eumelanins – A Possible Explanation for Broad Band Absorbance

M. Linh Tran[1†], Ben J. Powell[2] and Paul Meredith[1*]

[1]*Soft Condensed Matter Physics Group, School of Physical Sciences, University of Queensland, Brisbane, QLD4072, Australia*

[2]*Theoretical Condensed Matter Physics Group, School of Physical Sciences, University of Queensland, Brisbane, QLD4072, Australia*

[†]*Author now at: School of Chemistry, University of Bristol, Cantock's Close, Bristol, BS8 1TS, United Kingdom*

[*]*Author to whom correspondences should be addressed*

**Abstract**

We report the results of an experimental and theoretical study of the electronic and structural properties of a key eumelanin precursor – 5,6,-dihydroxyindole-2-carboxylic acid (DHICA) and its dimeric forms. We have used optical spectroscopy to follow the oxidative polymerization of DHICA to eumelanin, and observe red shifting and broadening of the absorption spectrum as the reaction proceeds. First principles density functional theory calculations indicate that DHICA oligomers (possible reaction products of oxidative polymerization) have red shifted HOMO-LUMO gaps with respect to the monomer. Furthermore, different bonding configurations (leading to oligomers with different structures) produce a range of gaps. These experimental and theoretical results lend support to the chemical disorder model where the broad band monotonic absorption characteristic of all melanins is a consequence of the superposition of a large number of inhomogeneously broadened Gaussian transitions associated with each of the components of a melanin ensemble. These results suggest that the traditional model of eumelanin as an amorphous organic semiconductor is not required to explain its optical properties, and should be thoroughly re-examined. These results have significant implications for our understanding of the physics, chemistry and biological function of these important biological macromolecules. Indeed, one may speculate that the robust functionality of melanins *in vitro* is a direct consequence of its heterogeneity, i.e. chemical disorder is a "low cost" natural resource in these systems.

**Keywords for Indexing**
5,6-dihydroxyindole-2-carboxylic acid
DHICA
5,6-dihydroxyindole
DHI
Melanin
Eumelanin
Polymerisation
UV-visible spectroscopy

# INTRODUCTION

The melanins are a broad class of functional macromolecule found throughout nature (1). In humans they serve as our primary photoprotectant in the hair, skin and eyes, but they are also found in the inner ear and brain stem, where their roles are unclear (2). Pheomelanin (a brown-red pigment derived from cysteinyl-dopa) and eumelanin (a brown-black pigment formed from the oxidative polymerization of dihydroxyindolequinones) are the predominant forms in humans. Eumelanin is the more prevalent melanin pigment and has long been thought to be synthesized *in vitro* from tyrosine via the Raper-Mason enzymatic pathway (3,4). To date, biochemical and biophysical research has primarily focused on understanding the basic photochemistry, photobiology and photophysics of these important substances, and has been stimulated by the key role that melanins play in photoprotection and their potential involvement in the development of melanoma skin cancer (2). In addition to this biologically directed research, melanins have also attracted interest from molecular biophysicists (5-7), quantum chemists (8-15), and more recently, the functional materials and condensed matter physics communities (16) due to their rather unique physio-chemical properties. Melanins exhibit broad band absorbance in the UV and visible (Figure 1), and conduct electricity in the condensed-phase (17,18). They have also been shown to photoconduct (19), and have a strong relative humidity dependent conductivity at room temperature (20). Additionally, and consistent with its role as a photoprotectant, eumelanin has been shown to efficiently de-activate UV and visible photon energy, with radiative quantum yields of <0.05% (21). Such properties have led to the proposition that melanins may be of some use as bio-mimetic functional soft solids (16,22) in applications such as photovoltaics, gas sensing and photothermal detectors. The prevailing paradigm is that melanins (eumelanin in particular) are condensed-phase organic semi-conductors, and that the broad band absorption (and indeed other electronic properties) can be explained by invoking a band structure model characteristic of an amorphous solid (17,23).

Despite extensive experimental studies conducted on both natural and synthetic melanins, the structure, composition and aggregation behaviour of these systems are not well understood. This is in part due to their adverse chemical and physical properties such as low solubility in most common solvents (5,24), difficulty in separation by chromatography, and opacity. At the primary structural level, it is fairly well accepted that eumelanins (the brown-black type of melanin, and the system of interest in this paper) are macromolecules of the various redox forms of 5,6-dihydroxyindole-2-carboxylic acid (DHICA, **1a**) and/or 5,6-dihydroxyindole (DHI, **1b**) (Figure 2) (1). These redox forms include dihydroxyindole (DHI), semiquinone (SQ) and indolequinone (IQ) species (10,14). However, opinions as to the appropriate model at the secondary structural level remain divided. (Note, in this paper we will use the phrase "secondary structure" to refer to the supramolecular assembly of monomers in a nomenclature borrowed from proteins.) Historically, eumelanin has been viewed as a highly conjugated, spatially extended homo or heteropolymer (23,25). However, in recent times, based primarily on a small number of X-ray scattering experiments (26) and attendant simulations, it has been proposed that eumelanin consists of small oligomeric units (<10 monomers) condensed by π-stacking into 4 or 5 oligomer nano-aggregates (26-29). This proposition requires a nanometer-sized protomolecule (the secondary structure) composed of sheets of covalently bonded DHI/DHICA monomers (with lateral extents of 15-20 Å). These protomolecules might then be stacked with ~3.4 Å intersheet spacing in a "graphitic-like structure" to form the aggregate. There remains some debate as to which model is appropriate.

Computational tools (*ab initio* and semi-empirical) have also be used to shed light on this structural question. Following the pioneering studies of Galvao and Caldas (10), notable works by Bolivar-Martinez *et al.* (11), Bochenek and Gudowska-Nowak (12), Ill'ichev and Simon (13), Stark *et al.* (14,15) and the current authors (8,9) have attempted to provide corroborative evidence for the oligomeric model by calculating the structure and electronic properties of DHI monomers and associated small (2-6 monomer) oligomers. Our calculations using density functional theory (DFT) on both DHI and DHICA have shown that different redox states possess different HOMO-LUMO gaps (the energy difference between the highest occupied molecular orbital and the lowest unoccupied molecular orbital) (8,9). Several authors (8,9,12,14,15) have suggested that the broad absorption may be due to the overlapping absorption features of a large number of individual, chemically distinct species that constitute the heterogeneous eumelanin ensemble. This model, based upon what we term as "chemical disorder" is an intriguing proposition (30), but, along with the oligomeric structural model, requires experimental proof.

Natural melanin is comprised of a combination of DHI and DHICA with great variability in the amount of these two precursors in the macroscopic pigment (3,4,31). In this context, it is useful to prepare synthetically pure DHI and DHICA monomers and monitor the formation of larger molecules as oxidative polymerization proceeds. The carboxylated form of the monomer is significantly easier to work with since DHI tends to spontaneously react under even the mildest conditions. In the current study we use DHICA as our model "eumelanin precursor". We report its synthesis and detailed chemical characterization, and use UV-visible spectroscopy and DFT as tools to understand how the electronic structure of the reaction products evolve during oxidative polymerization. Our data suggests that an amorphous organic semi-conductor model is not required to explain the broad band absorbance of eumelanin, i.e. our data is consistent with the "chemical disorder" model. We believe it to be the first experimental evidence in support of this "chemical disorder" theory.

## MATERIALS AND METHODS

**Materials**

DL-3,4-dihydroxyphenylalanine (DL-dopa) and potassium ferricyanide were purchased from Sigma Aldrich Pty. Ltd (Sydney, Australia). All other chemicals were of analytical grade. Milli Q water (resistance = 18 MΩ.cm) was used throughout the experiments.

**Synthesis of 5,6-Dihydroxyindole-2-carboxylic acid (DHICA)**

The synthesis of DHICA was adapted from a procedure reported in the literature (32). Potassium ferricyanide solution (7 g, 3.04 mmol) in 60 mL water was added dropwise to a stirred solution of DL-dopa (1 g, 5.07 mmol) in 500 mL of water under a nitrogen atmosphere in a glovebox for 5 minutes and then the mixture was allowed to stir for 8 minutes. A saturated solution of sodium sulfate was then added to the mixture to terminate the reaction and the solution was allowed to stir for a further 8 minutes. The mixture was acidified to pH 2 using 6 M HCl and stirred for 10 minutes. The product was then extracted in ethyl acetate (~500 mL). The ethyl acetate portion was then washed with saturated sodium chloride solution twice, followed by a single wash with water. While under nitrogen, the ethyl acetate extract was transferred to a rotary evaporator and the solution was concentrated under vacuum. The reaction flask was then filled with nitrogen and transferred back into the glovebox. Hexane was added to precipitate the product, which was collected by vacuum filtration. The product was purple-grey in colour (this is impurity related colour): m.p 220°C; EI-MS: m/z calculated for $C_9H_7NO_4$ ($M^+$) 193.15, found 193; Anal. Calc. for $C_9H_7NO_4$: C, 53.95; H, 3.85; N, 6.53, found: C, 53.81; H, 3.87; N, 6.53%. $\lambda_{max}$ (DMSO) = 320 nm.

**UV-visible Kinetic Experiments**

The oxidation of DHICA in slightly basic (pH ~ 9) solution was followed using a Perkin Elmer Lamda λ40 UV-visible spectrometer. Absorption spectra between 200 – 600 nm were recorded at a scan rate of 240 nm/min using a slit width of 4 nm and spectral resolution of 4 nm. Spectra were collected using a quartz 1 cm cuvette. Aliquots from a 2 mM solution were examined at time intervals of 0, 0.5, 1, 2, 4, 8, 12, 24, 51, 74 hrs.

**Instrumentation (NMR & X-ray Photoelectron Spectroscopy (XPS))**

$^1$H and $^{13}$C NMR spectra were run using a Bruker DRX-500 high-resolution NMR spectrometer interfaced to a 11.7 Tesla 51 mm bore magnet system. DMSO-d6 was used as the solvent. XPS spectra were acquired with a Kratos Axis ULTRA X-ray Photoelectron Spectrometer incorporating a 165 nm hemispherical electron energy analyzer. The incident radiation was monochromatic Al $K_\alpha$ X-rays (1486.6 eV) at 150 W (15 kV, 10 mA). Survey (wide) scans were taken in the range of 0 – 1100 eV at a pass energy of 160 eV and multiplex (narrow) high resolution scans were taken with a pass energy of 20 eV. Base pressure in the analysis chamber was $10^{-9}$ Torr during sample analysis. The DHICA powder was packed thickly onto a piece of black carbon tape.

The XPS spectral envelopes were resolved into component peaks using Gaussian-Lorentzian (10 - 20 % Lorentzian) curves. The binding energy (BE) of the spectra were referenced by assigning the main C$1s$ hydrocarbon peak to 284.7 eV (33).

**Calculation Details**

The chemical and electronic structures were calculated from first principles DFT. We performed our calculations using the Naval Research Laboratory Molecular Orbital Library (NRLMOL) (34-40). NRLMOL performs massively parallel electronic structure calculation using Gaussian orbital methods. Throughout we have used the Perdew, Burke and Ernzerhof (PBE) (41) exchange correlation functional, which is a generalised gradient approximation (GGA) containing no free parameters. For each molecule we have fully relaxed the geometry with no symmetry constraints.

To calculate the structure of a molecule within DFT two nested problems must be solved self-consistently. Firstly, the self consistent field (SCF) problem, which amounts to solving the Kohn-Sham equations within the Born-Oppenheimer approximation for a fixed set of nuclear coordinates. Secondly, the Hellman-Feynman theorem is used to calculate the forces on the nuclei and thus iteratively search for a local minimum of the energy in the space of structures. In any self consistent solution criterion must be chosen to allow one to decide when the problem is converged. In this work we consider the SCF problem to be converged when successive iterations differ in total energy by less than $10^{-8}$ Ha and geometries to be converged when the largest force acting on any atom are less than $10^{-3}$ Ha/Bohr.

In all of the calculations presented in this paper we use Porezag-Pederson (PP) basis sets (42), which, in contrast to standard approaches, are not a fixed set of basis functions, but are adjusted based on a total-energy criterion. PP basis sets have been proven to satisfy the $Z^{10/3}$ theorem and, for weakly interacting atoms, it has been shown that PP basis sets have no superposition error (43). These basis sets are available on request.

DFT is a theory of the ground state; therefore calculations represent the energy gap between Kohn-Sham eigenvalues and not the true HOMO-LUMO gap of the molecules. This is known as the band gap problem (44). Therefore we have also employed the ΔSCF method (44) to calculate the HOMO-LUMO gap. We have previously shown that equivalent results for DHI reproduce the trends found in time dependent DFT calculations (8).

## RESULTS & DISCUSSION

**Characterisation of 5,6-dihydroxyindole-2-carboxylic acid (DHICA)**

Several single-step synthetic approaches to DHICA have been published (32,45,46). For the research reported here, a modification of Kroesche's method was used. The $^1$H NMR spectrum of DHICA (Figure 3) revealed broad resonances at 8.59 and 9.10 ppm associated with the hydroxy groups in the 5 and 6 positions. Another important feature is the 11.13 ppm peak assigned to the indole amine proton. Finally the carboxylic acid proton broad band appears at 12.31 ppm. The other resonances corresponding to the aromatic ring protons at 6.89, 6.85 and 6.77 ppm also appear. The values for the $^1$H NMR peaks are in close agreement with literature reports (32,47). The assignments for the $^{13}$C NMR spectrum are shown in Figure 4. Of significance are the signals for the –COOH moiety at 162.75 ppm and the –OH groups at 146.11 and 141.98 ppm. There was no observation of the presence of quinone in the $^{13}$C NMR spectrum. To further establish that the hydroxy groups did not oxidize into quinone groups, the XPS spectra of DHICA were acquired and interpreted. The C *1s* spectral envelope contained components arising from the various functional groups found in DHICA (Figure 5a). The assignments for the components are presented in Table 2. The COOH group occurs at 288.7 eV, a binding energy shift of 4 eV from the main aromatic signal. The aromatic shake-up peaks can also be observed at higher binding energies to the COOH band. The only other peaks present were those due to the C-N or C-OH atoms and no C=O functionality was observed suggesting conversion to the quinone form did not occur. The N *1s* and O *1s* spectra (Figure 5b and 4c) provided corroborative evidence for the structure of DHICA. It was observed that only one type of nitrogen signal arose and it was assigned to the indole nitrogen group. In the O *1s* spectrum the area ratio of C-OH and C=O peaks at 533.53 and 531.86 eV, respectively (33), agreed with theoretical predictions of 3:1. The NMR and XPS spectra, along with the mass spectral information and microanalysis results reported in the experimental section provide comprehensive characterization of the melanin precursor.

**UV-Visible experiments**

The evolution of the DHICA monomer into larger "eumelanin" molecules was followed using UV-visible spectroscopy. Figure 6a shows the absorption spectrum of a weakly basic DHICA solution at the start of this experiment. Initially one prominent absorption peak appears centered at 328 nm corresponding to the lowest energy DHICA transition which calculations show to be a π−π* transition (note that there is a node in the electron density in the plane of the molecule of the HOMOs and LUMOs reported in reference 9). Taking the HOMO-LUMO gap to correspond to the lowest energy transition, an experimental value for this gap of 3.8 eV can be extracted (which is consistent with the HOMO-LUMO gap of 3.0 eV calculated for a DHICA monomer in reference 9, given the band gap problem discussed above and the fact that calculations neglect solvent effects (47), see below). This suggests that at this stage only the monomer is present in solution. Over time, the absorption maximum red shifts and the initial peak broadens (Figure 6b) in a progressive fashion until it almost disappears at the end of the monitoring time (74 hours). This shift is likely to be a convolution effect as absorption by reaction products (higher order species) at lower wavelengths smear out the initial 328 nm absorption maximum. The dramatic broadening of the absorption profile also offers evidence for the formation of reaction products with lower energy HOMO-LUMO gaps (see the dimer results in the next section). At the end stages, the absorption profile, although never becoming monotonic, approaches the broad spectrum characteristic of eumelanin solutions (49). It has previously been suggested that this broad absorbance observed in all melanins can be explained in terms of structural and chemical heterogeneity (30). In this "chemical disorder" model, melanin solutions (or indeed condensed phase films or powders) are ensembles containing a large range of chemically distinct macromolecules (oligomeric or polymeric), each with a different HOMO-LUMO gap. The exponential monotonic absorption is then a result of the superposition of a large number of inhomogeneously broadened Gaussian transitions associated with each of the components of the ensemble. We have previously predicted that DHICA macromolecular ensembles contain less "effective chemical disorder" than the equivalent DHI macromolecular ensembles (8,9). This prediction is based upon the fact that the (oxidised) quinone forms of the DHICA monomer are energetically unstable and therefore not likely to form. This is not the case for DHI. The fact that the oxidative polymerization of DHICA does not lead to a truly monotonic profile is entirely consistent with this prediction, as is the fact that we do not see a quinone in the analysis of the monomer.

It has been suggested that scattering contributes significantly to measurements of melanin absorption (50). A broadening of the spectrum into the visible would require Mie scattering from un-dissolved particles in solution. Several authors have shown that scattering is significant under specific conditions of neutral or low pH and high concentration, i.e. in the low solubility limit (51,52). The eumelanin solutions can even appear turbid and milky. In the study by Vitkin et al. it was estimated that scattering contributed a maximum of 20% to the absorption coefficient. However, there is now a significant body of literature to confirm that the monotonically decreasing absorption with increasing wavelength in melanins is in fact due to "real absorption" (see for example 53,54) – especially under the types of conditions we employ in our study: alkaline pH and low concentrations. We have also recently shown that working within this regime in combination with a mathematical procedure to account for excitation attenuation and emission re-absorption allows one to extract quantitative information from optical emission and excitation spectra of eumelanins (21,55).

The absorptions at 328 nm and 250 nm were examined more closely (Figure 7). It was found the absorption by the monomer precursor decreased linearly as a function of time. This result suggests that the consumption of DHICA molecules is constant. The absorption band at 250 nm progressively increases as the reaction proceeds. It is conceivable that this feature is associated with transitions to higher lying states in reaction products. The growth of these products is non-linear as a function of time, and was found to obey the relation:

$$conc. = \frac{b + ct}{1 + at} \qquad (1)$$

where b, c and a are constants. Fitting the data to the equation 1, the extracted values for the constant are as follows: a = 0.152 hr$^{-1}$, b = 1 mol ltr$^{-1}$, c=0.25 mol ltr$^{-1}$hr$^{-1}$. In general, such complex kinetics indicates the presence of competing forward and backward reactions, whose rates are dependent upon the availability of precursor. It would also indicate that the reaction proceeds via a stable intermediate (for example a small oligomeric species) which ceases to be the primary reaction product at a critical concentration of available reactant, and becomes the building block for subsequent products. This idea of an oligomeric "proto-molecule" is consistent with a heterogeneous ensemble picture.

**Quantum Chemical Calculations**

It has previously been shown (9) that, *in vacuo*, DHICA is significantly more stable than its oxidized forms (the quinone and semiquinone forms). We therefore restrict our attention to the 5,6-dihydroxy form, which is labeled as **1a** in Figure 2. Also shown in Figure 2 are four possible dimers of DHICA (**2a**-**2d**). We study these molecules as a representative sample of the type of molecules that will form as a result of oxidative polymerization. Clearly this is not the complete set of all possible dimers, but our objective here is to examine the implications of polymerization with respect to the HOMO-LUMO gaps of the products. In Table 3 we report the total energy calculated for molecules **2a**-**2d** relative to twice that of **1a**. We also calculate the relative abundances of these dimers at 300 K (Table 3), neglecting entropic effects. Here something of puzzle emerges. Our calculations indicate that none of the dimers we have considered are stable with respect to two monomers in reactions such as **2a** + 2$H_2$ → 2(**1a**), i.e. they should not form. However, it is by no means clear that this represents the situation in the actual reaction mix. In particular, it is important to note that our calculations represent the situation *in vacuo*. The stability of the dimers may be somewhat different in a solvent - in particular the hydrogens given off in the reaction are unlikely to form $H_2$ (which indicates that we have significantly underestimated the enthalpy of the dimers relative to the monomer) and further solvent effects may significantly alter the relative enthalpy and, perhaps more significantly, the entropy associated with each molecule. For example, it has been previously predicted that significant hydrogen bonded networks are formed around DHI monomers in an aqueous environment (8). It should also be noted that several stable dimers are known for DHI (reference 8 and references therein). We have therefore also reported the energy of vaporization per atom in Table 3. In contrast to the simple interpretation of the relative energies this suggests that the dimers with two inter-monomer bonds (**2a** and **2d**) are stable with respect to those with a single inter-monomer bond (**2b** and **2c**), and that all of the dimers are stable with respect to the monomer. The claim that dimers with two inter-monomer bonds are more stable than those with a single cross link is consistent with the results of Stark *et al.* (14) who found that planarity had significant effect in stabilizing dimers formed from IQ. However, the safest interpretation of the data presented in Table 3 is that it indicates that **2b** is stable relative to **2c** and **2d** is stable with respect to **2a**, but that no conclusions can be safely drawn of the relative stabilities of the dimers and monomers. However, it is clear from figures 6 and 7 that the reactions moving from the monomer to the macromolecule is extremely slow, which may be what the apparent low enthalpy of the dimers is reflecting.

In Table 4 we compare the HOMO-LUMO gap found from a simple interpretation of the Kohn-Sham eigenvalues with those found by the ΔSCF method for molecules **2a**-**2d**. For comparison we reproduce the previously published (9) result for **1a** in Table 4. Once again, we note that these calculations represent the *in vacuo* situation, and one would expect the presence of a polar solvent to inhomogeneously broaden and shift these results (whether this is a red shift or a blue shift will depend, in general, on the details of the molecule-solvent interaction – but the direction and relative magnitude of the shift would be consistent for all molecules) (45). Despite this caveat, it is clear that the HOMO-LUMO gaps of all four dimers are significantly red shifted with respect to the monomer, i.e. even the lowest molecular weight reaction products will have absorption maxima at longer wavelengths than the original precursor (DHICA). One can clearly see why this red shifting occurs by examining the electronic densities of the HOMOs and LUMOs of molecules **2a**-**2d** (figure 8). Significant electron density is distributed spatially over the entire molecule in all of the orbitals. This extended delocalization naturally corresponds to a reduction of the energy gap relative to the parent monomer. It can also be seen that **2a** and **2d**, are further red shifted than **2b** and **2c** (c.f. Table 4). This appears to be a direct result of the greater delocalization afforded by the fact that **2a** and **2d** have two inter-monomer bonds rather than the single inter-monomer bond found in **2b** and **2c**. This result is also consistent with the findings of Stark *et al.* who found that planarity led to a reduction of the HOMO-LUMO gap in dimers of SQ (14). Furthermore, each dimer has a different gap value over a range of 0.2 eV. These are important results since they confirm that a simple ensemble containing the monomer and a relatively small number of dimers would indeed produce an absorption profile approaching that observed for the polymerized DHICA solutions. Based upon this analysis, widening the ensemble to include other monomer types (DHI and its oxidized forms as in real eumelanins), a greater range of dimers, and higher order oligomers (trimers, tetramers, etc. with further red shifted HOMO-LUMO gaps) all with different abundances and absorption cross-sections, would be expected to produce the characteristic melanin broad band monotonic absorption. Hence, the traditional extended homopolymer or heteropolymer organic semiconductor model is not required to explain the optical absorption of eumelanin. The secret to the robust functionality of these mysterious functional macromolecules may be associated with the extreme chemical disorder inherent in their secondary structure. If true, this new structural model has profound implications for our understanding of the physics and chemistry of melanins. In particular, the long held view that they are condensed phase semiconductors should be re-examined. Additionally, from the biological function perspective (melanins are the primary photoprotectants in humans) we should consider how these ensembles are capable of nearly complete deactivation of UV-visible photon energy. Finally, in a more generic sense, it may be interesting to speculate whether the rather "low cost" resource of chemical disorder is used elsewhere in nature to create robust electronic functionality (for example, coupling heterogeneity in the antenna chlorophylls of PSI of cyanobacteria is thought to red shift and broaden the antenna absorption cross-section (56)), or whether the melanins are a unique case.

**CONCLUSIONS**

In this paper, we have reported the synthesis of a stable and convenient eumelanin precursor, namely 5,6-dihydroxyindole-2-carboxylic acid. This moiety is present in the natural and synthetic forms of eumelanin together with DHI and its reduced forms. We have used UV-visible spectroscopy to follow the oxidative polymerization of DHICA, and observe red shifting and broadening of the absorption spectrum as the reaction proceeds. First principles density functional theory calculations indicate that DHICA oligomers (the likely reaction products of oxidative polymerization) have red shifted HOMO-LUMO gaps with respect to the monomer. Furthermore, different bonding configurations (leading to oligomers with different secondary structures) produce a range of gaps. These experimental and theoretical results lend support to the chemical disorder model where the broad band monotonic absorption is a consequence of the superposition of a large number of inhomogeneously broadened Gaussian transitions associated with each of the components of the eumelanin ensemble. We therefore conclude that the traditional model of melanin as an amorphous organic semiconductor is not required to explain its optical properties, and should be thoroughly re-examined.

## Acknowledgements

We would like to thank Tunna Baruah, Noam Bernstein, Paul Burn, Joel Gilmore, Ross McKenzie, Mark Pederson, Jenny Riesz, Jeff Riemers and Tad Sarna for enlightening discussions. BJP would like to express his gratitude to Ross McKenzie for financial support during this research. The work was funded by the Australian Research Council (Discovery Program), and calculations were performed on the Australian Partnership for Advanced Computing (APAC) National Facility under a grant from the Merit Allocation Scheme.

| Peak / ppm | Assignment | Literature |
| --- | --- | --- |
| 96.92 | C7 | 97.29 |
| 104.87 | C3 | 105.27 |
| 107.02 | C4 | 107.43 |
| 119.86 | C8 | 120.22 |
| 125.84 | C2 | 126.13 |
| 132.60 | C9 | 132.97 |
| 141.11 | C6 | 142.32 |
| 146.11 | C5 | 146.46 |
| 162.75 | COOH | 163.03 |

**Table 1:** $^{13}$C NMR of DHICA in DMSO-d6, peak assignments (relative to those presented in reference 44).

| BE / eV | Assignment |
|---------|------------|
| 284.69  | C aromatics |
| 285.55  | C-N |
| 286.26  | C-O |
| 288.66  | C=O |

**Table 2:** XPS carbon *1s* peak positions for DHICA (from Figure 5a).

| Molecule | Total Energy (eV) | Relative Abundance *in vacuo* at 300 K | Energy of vaporization per atom (eV) |
|---|---|---|---|
| **2a** | 3.82 | $<10^{-32}$ | -1004.157 |
| **2b** | 1.96 | 1 | -957.511 |
| **2c** | 4.02 | $<10^{-34}$ | -954.738 |
| **2d** | 2.89 | $<10^{-15}$ | -1004.181 |

**Table 3:** Total energy (eV) and relative abundance of four possible dimers of DHICA (cf. Figure 2). The total energies are quoted relative to twice the energy of 1a and are adjusted for chemical change by assuming that all of the removed hydrogens form $H_2$. The relative abundance is calculated assuming a Boltzmann distribution at 300 K. This naïve treatment of the calculated energies is probably inadequate. In particular, the energy of the hydrogen is likely to depend on the detailed chemistry of the solvent. Indeed this probably means that it is not correct to compare the energies of 2a and 2d with those of 2b and 2c. Another way to estimate the relative stability of the molecules is to calculate the energy required to vaporize the molecule per atom. This is therefore also presented in the table. It can be seen that significantly more energy per atom is required to vaporize 2a and 2d than 2b or 2c, suggesting that the former are more stable. (Note that the equivalent figure for molecule 1a is 910.126 eV, suggesting that the dimers are, in fact, stable). However, it is clear that 2d is significantly more stable than 2a and that 2b is significantly more stable than 2c.

| Molecule | Simple interpretation of the Kohn-Sham eigenvalues | ΔSCF |
| --- | --- | --- |
| **1a** | 2.85 eV | 3.04 eV (407 nm) |
| **2a** | 2.02 eV | 2.13 eV (581 nm) |
| **2b** | 2.24 eV | 2.32 eV (534 nm) |
| **2c** | 2.28 eV | 2.46 eV (504 nm) |
| **2d** | 2.04 eV | 2.26 eV (549 nm) |

**Table 4:** The HOMO-LUMO gap (in eV) for the monomer (1a) (reproduced from Ref. 23) and four possible dimers (2a-2d). We compare the value found from the simple interpretation of the Kohn-Sham value, which tends to underestimate the HOMO-LUMO gap because of the band gap problem, with the results of ΔSCF calculations, which are known to much more accurate (41). It is previously been shown (8) that, for DHI, ΔSCF calculations are in good agreement time dependent DFT calculations. All of the dimers show a significant red-shift compared to the monomer.

**Figure 1:** The characteristic broad band absorbance of eumelanin and pheomelanin. The spectra are featureless from the UV to the NIR, and are more characteristic of an inorganic semiconductor than an organic pigment.

**Figure 2**: Schematic representation of 5,6-dihydroxyindole-2-carboxylic acid (DHICA) (1a), 5,6-dihydroxyindole (DHI) (1b) and the four dimers (2a-2d) which we also consider here.

**Figure 3:** $^1$H NMR of DHICA in DMSO-d6.

**Figure 4:** $^{13}$C NMR of DHICA in DMSO-d6.

**Figure 5:** XPS spectra of DHICA a) C 1s region b) O 1s region c) N 1s region. See text and Table 2 for assignments of individual components.

**Figure 6:** UV-visible absorption spectra DHICA in NaOH solution at t = 0 a) and at various times b) (legend represents hours).

**Figure 7:** Absorption versus time plots of the DHICA solution at 328 nm and 250 nm.

**Figure 8:** The electron density in the highest occupied molecular orbital (HOMO, left) and lowest unoccupied molecular orbital (LUMO, right) of 2a-d.

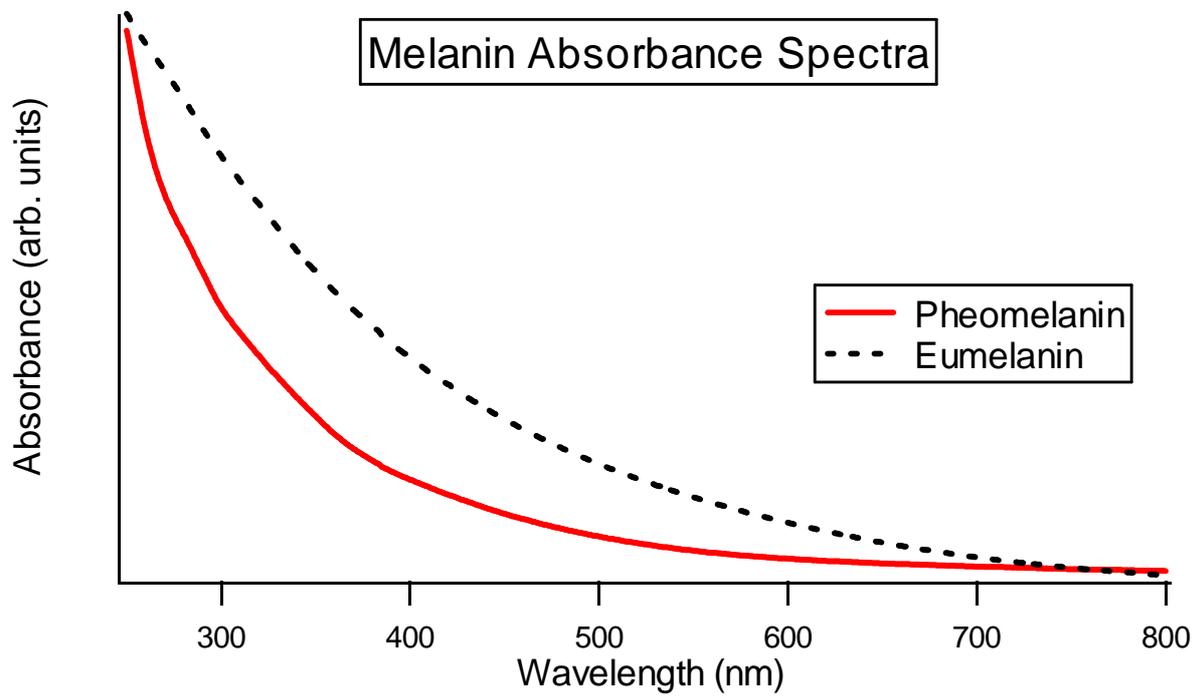

**Figure 1**

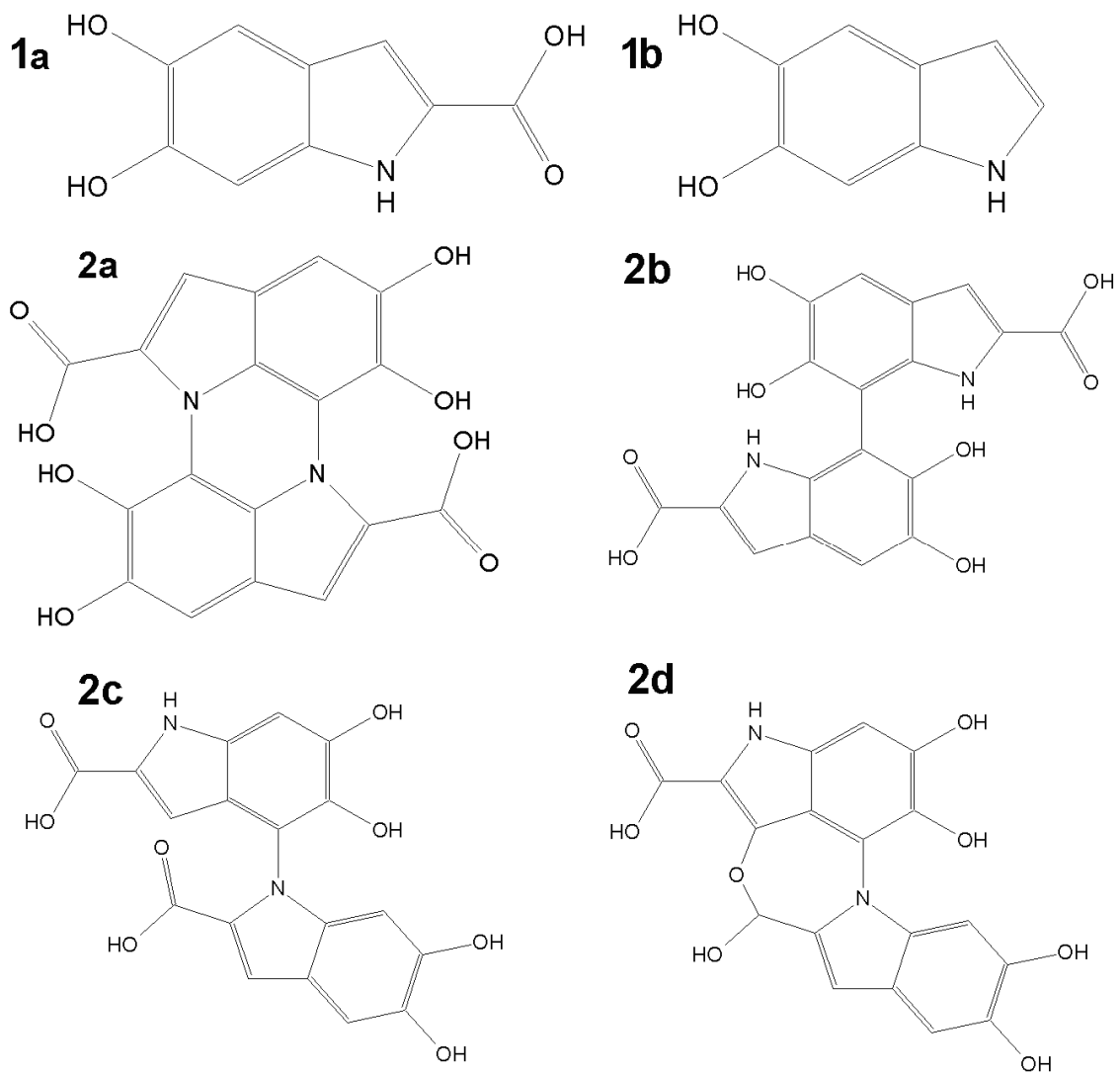

Figure 2

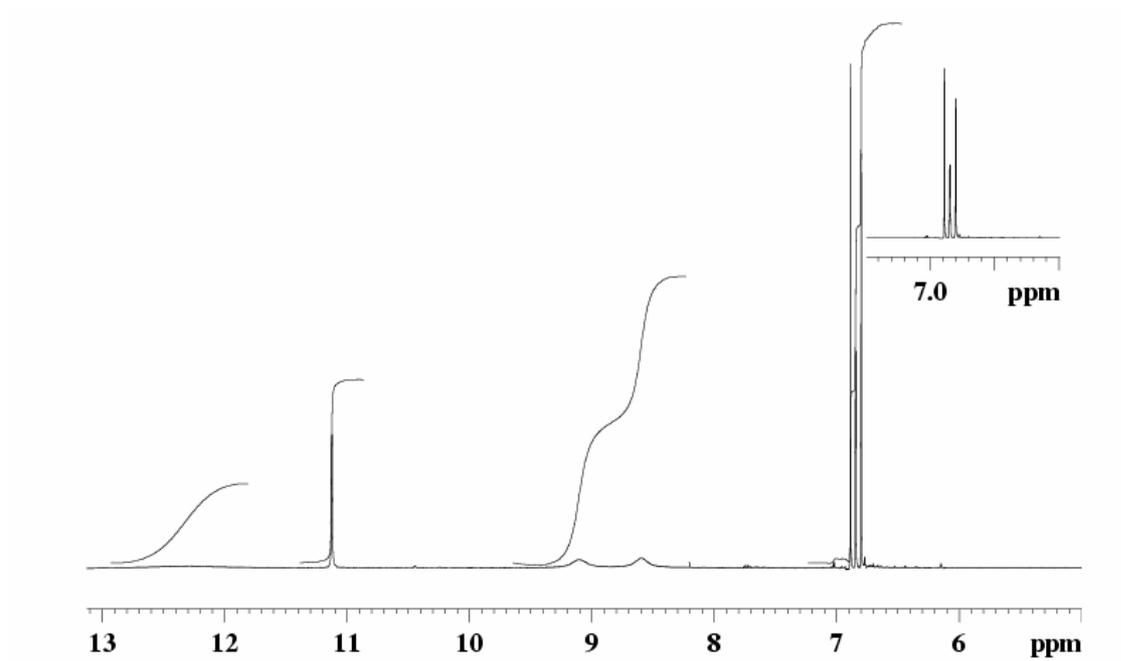

**Figure 3**

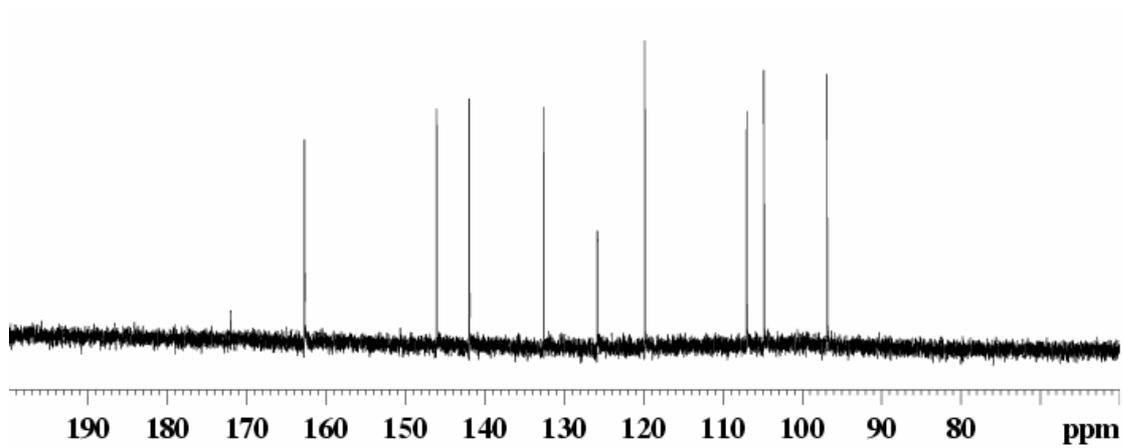

**Figure 4**

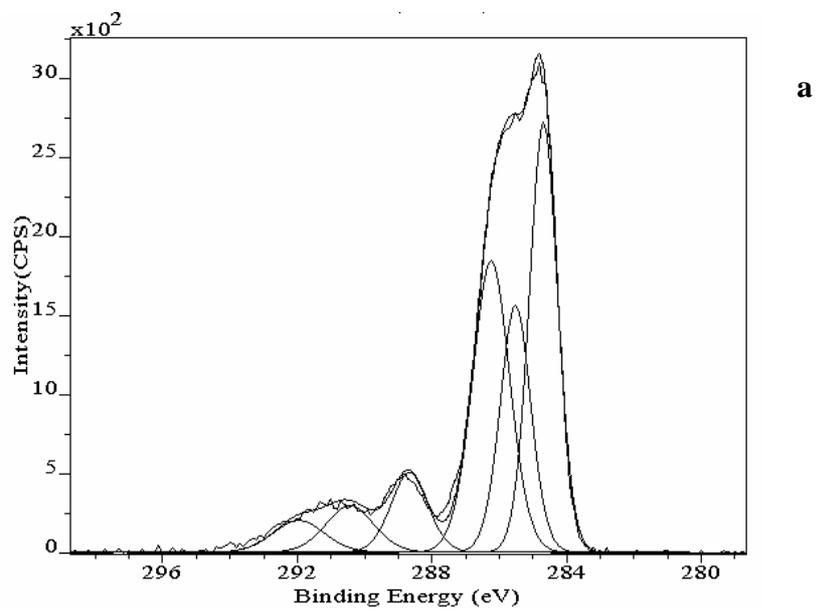

**Figure 5a**

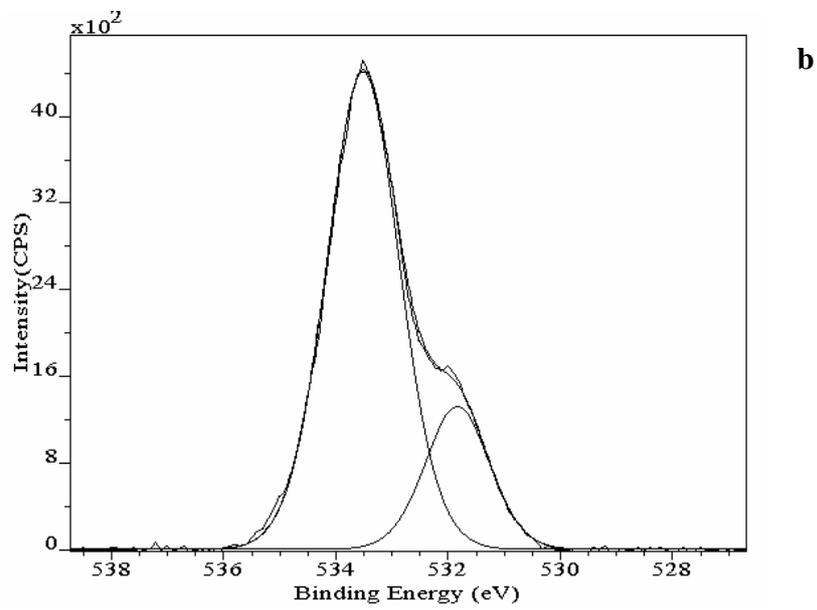

**Figure 5b**

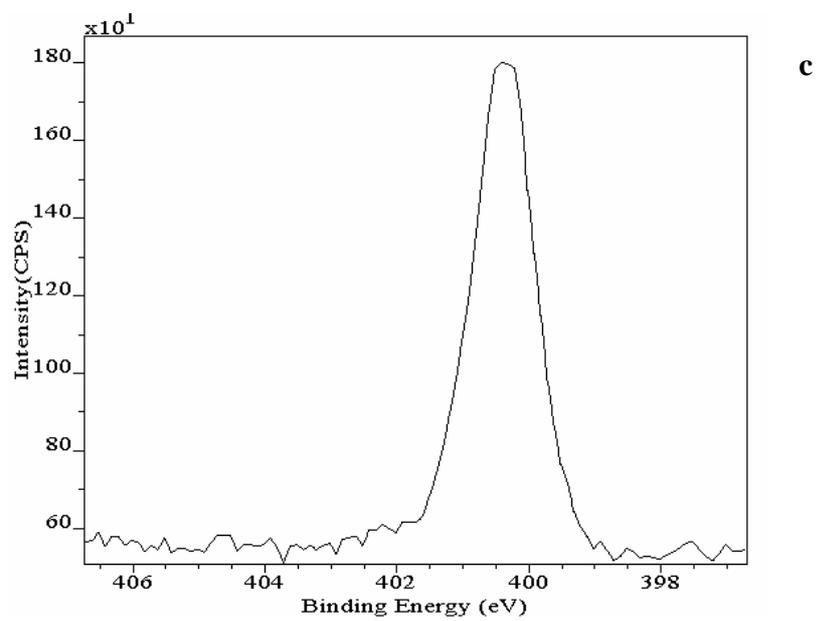

Figure 5c

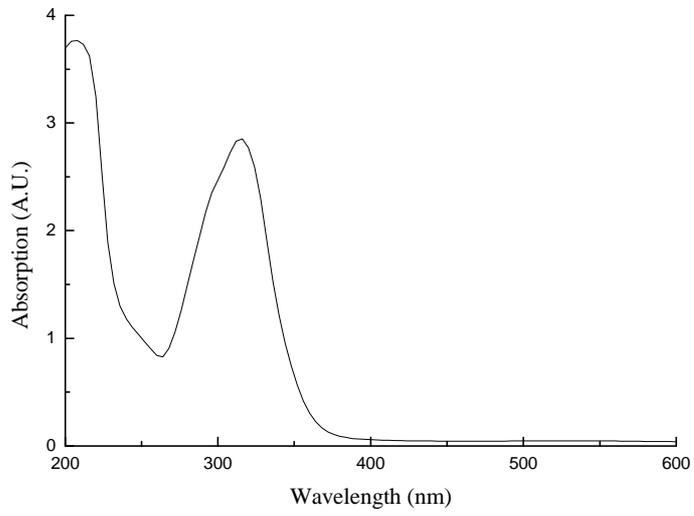

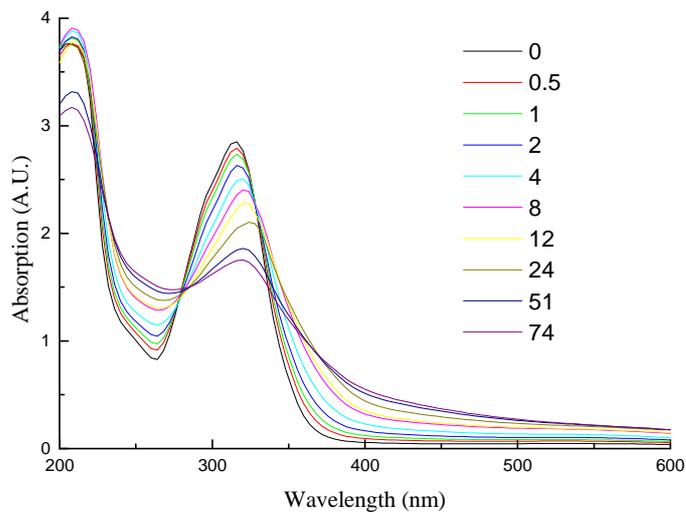

**Figure 6**

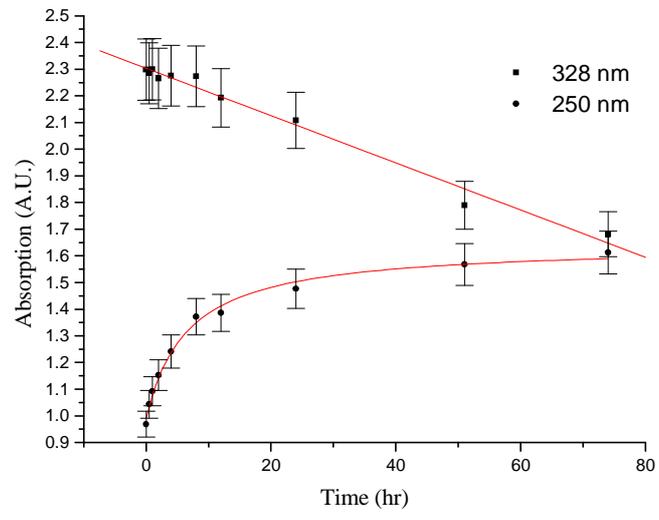

**Figure 7**

HOMO   LUMO

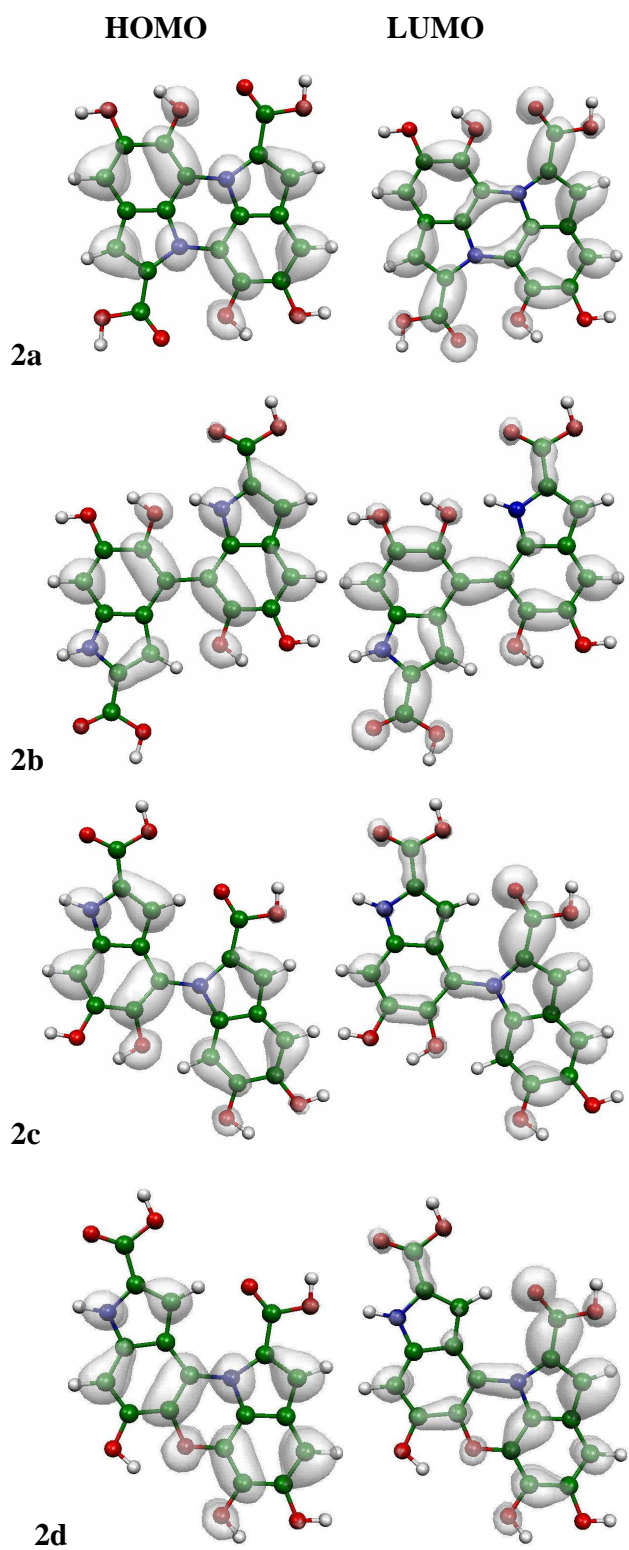

**2a**

**2b**

**2c**

**2d**

**Figure 8**